\documentstyle[aps,epsfig]{revtex}                         

\draft
\begin{document}
\twocolumn[\hsize\textwidth\columnwidth\hsize
     \csname @twocolumnfalse\endcsname



\title{Numerical simulation of  Einstein-Podolsky-Rosen experiments
       in a local hidden variables model.
       }

\author{W. A. Hofer}
\address{
         Department of Physics and Astronomy,
         University College, Gower Street, London WC1E 6BT, UK}

\maketitle

\begin{abstract}
We simulate correlation measurements of entangled photons
numerically. The model employed is strictly local. In our model
correlations arise from a phase, connecting the electromagnetic
fields of the two photons at their separate points of measurement.
We sum up coincidences for each pair individually and model the
operation of a polarizer beam splitter numerically. The results
thus obtained differ substantially from the classical results. In
addition, we analyze the effects of decoherence and non-ideal beam
splitters. It is shown that under realistic experimental
conditions the Bell inequalities are violated by more than 30
standard deviations.
\end{abstract}

\pacs{03.65.Ud,03.65.Ta}

\vskip2pc]


Since experimental tests of the Einstein-Podolsky-Rosen (EPR)
problem became feasible
\cite{epr35,bohm51,freedman72,aspect82,weihs98}, the field has
gained a popularity and importance, which is startling. The
experiments themselves do not seem to justify this broad interest.
The fact that polarization measurements of two separated photons
are connected, even if the measurements are in space-like
separation \cite{weihs98}, is in itself problematic, but hardly
accounts for the large number of publications devoted to the
problem every year. The importance of the EPR problem derives from
its impact on two seemingly separate fields. On the one hand, the
theoretical predictions impose a limit on theoretical concepts,
aimed at modifying quantum mechanics by hidden variables models
\cite{bohm52,bell64}. Here, the current understanding is that the
experiments contradict any local model \cite{clauser74,clauser78}.
On the other hand, entanglement of separated photons is essential
for many theoretical concepts in quantum computation and quantum
information \cite{qubit}. Here, the question whether quantum
mechanics provides the only suitable model for these experiments,
is essential, because qubits, the principal information units in
these concepts, are only defined within the quantum formalism.

Theoretically, the entanglement of separated photons rests on the
validity of the superposition principle. The question, whether
superposition is a sufficient condition for non-locality, is open.
Deutsch and Hayden showed recently that all information in an EPR
experiment is stored locally, even though it cannot locally be
accessed \cite{deutsch99}. In a different approach it was shown
that the observed connection can be interpreted as fourth-order
optical correlation. The connection then arises due to the phase
of the photons' electromagnetic fields
\cite{hofer00,kracklauer00}. All physical effects in this case are
local, even though the correlations must be expressed in terms of
polarizer settings at different locations.

In this paper we present a model, where the phase connection
between the two photons is retained, but the superposition
principle for the separate photon fields is discarded. Essentially
this means that interference terms in the correlations are
omitted. We shall show, however, that the model also in this case
is in accordance with experimental results
\cite{freedman72,aspect82,weihs98}, provided the polarizer beam
splitter (PBS) is not completely ideal. The non-ideal PBS in our
model can only discriminate between the horizontal and vertical
polarization of photons, if the intensities of the photon's
electromagnetic field components, projected onto the orthogonal
polarizer axes, differ by more than 30\%. We shall show that this
threshold is consistent with and necessary for the high visibility
observed in state-of-the-art experiments. Furthermore, we shall
establish that the Bell inequalities are violated under these
conditions by more than 30 standard deviations. Single photon
experiments will be treated within the same model in a separate
paper.

The setup in our model consists of a source for entangled photons
and two PBS with photon detectors. The source, at $z = 0$ of the
coordinate system, emits two entangled photons. Their planes of
polarization are unknown. But it is known that these planes at the
moment of emission are perpendicular to each other ($\triangle
\varphi = \pi/2$). The polarizers are at $z = \pm L$, where $L$ is
an integer multiple of the photons' wavelength $\lambda$. Since
laser emissions have finite extensions in the range of a few
microns \cite{photons}, we describe each photon as an excitation
of the electromagnetic field of limited extension. To simplify the
description these fields shall be monochromatic. That is to say we
disregard the effects of higher Fourier components, which arise
from the limited extension of the fields. Then the phase along the
photon's path of propagation is equal to the phase of a
monochromatic wave. For photons of circular polarization the angle
of polarization at a point $z_0$ depends on the distance from the
source and the initial angle of polarization. For linear
polarization only the amplitude of the field depends on the
distance $z_0$, but not the angle of polarization. This is
changed, however, in EPR experiments using linear polarized light,
by an optical modulator \cite{modulator}. A modulator shifts one
component of the electromagnetic field, either the horizontally or
the vertically polarized one by a phase-angle $\triangle \psi$. If
the field is initially in linear polarization, it is described by
$\vec{E}(z,t) = \vec{E}_0 \cos (kz - \omega t)$, where $\vec{E}_0
= (E_{0X}, E_{0Y})$. The angle of polarization $\varphi_0 =
\arctan \left(E_y / E_x \right)$ in this case is independent of
the phase. If one component, say $E_y$ is shifted by a phase-angle
$\triangle \psi$, then the field after the modulator will be:

\begin{equation}
\vec{E}' = (E_{0X} \cos (kz - \omega t), E_{0Y} \cos (kz - \omega
t + \triangle \psi))
\end{equation}

The angle of polarization $\varphi$ is then no longer constant,
but depends on the phase:

\begin{equation}
\varphi = \arctan  \left\{\frac{E_{0Y}}{E_{0X}} \left[\cos
\triangle \psi - \tan (kz - \omega t) \sin \triangle \psi \right]
\right\}
\end{equation}

Therefore in both cases the angle of polarization $\varphi$
depends on the location $z$ of a measurement. The measurements are
performed with polarizer beam splitters and two photon detectors
\cite{aspect82,weihs98}. This differs from using optical
polarizers, because the combination PBS and detector yields not a
continuous result, e.g. the field intensity after the polarizer,
but a discrete one. The photon is either detected at detector 1,
or at detector 2. In one case, it is said to have vertical
polarization (+), in the other horizontal polarization (-). The
representation of this experiment by a classical optical model is
not straightforward. In Furry's integral, for example, the
correlation is described by \cite{furry36},

\begin{equation} \label{eq003}
P (\alpha,\beta) = \int d \varphi \cos^2 (\varphi + \pi/2 -
\alpha) \cos^2 (\varphi - \beta)
\end{equation}

where $P(\alpha,\beta)$ is the probability of a coincidence,
$\varphi$ the unknown angle of polarization, and $\alpha$ and
$\beta$ are the polarizer settings. In this equation it is assumed
that a single measurement can be described by the product
$\cos^2(\varphi + \pi/2 - \alpha) \cos^2(\varphi - \beta)$. But
this is evidently not the right representation for the dichotomic
results obtained in the experiments. In our view it is partly this
error, which is responsible for the deviation between the
correlation curves obtained in the experiments, and the
correlation curves obtained from Eq. (\ref{eq003}). Analyzing the
measurements event by event only two results are obtained in the
experiments on either PBS: (+), if the angle of polarization
$\varphi$ is between $\pi/4$ and $\pi/2$, and (-), if it is
between $0$ and $\pi/4$. In our model we formalize this feature of
the experiments by a switch at $\pi/4$; numerically this is
accomplished by a selection criterion based on the value of
$cos^2(\varphi)$. The model in this case encounters a difficulty,
which is absent in a classical representation. It is the threshold
of the PBS at the critical angle of $\pi/4$. For polarization
angles close to this value the intensity in both channels of a PBS
is equal. Theoretically, this would mark an event, where both
polarizations, (+) {\it and} (-), are recorded for a single
photon. Experimentally, such a case does not exist. Therefore we
must impose a suitable threshold at this angle, to decide on the
actual outcome. In our view this threshold is an experimental
parameter and cannot be determined from theoretical estimates
alone. We shall therefore use the experimental results, in
particular the observed visibility, to determine this value.
Formally, this reduces the range of the model PBS to the values
$\cos^2(\varphi) \ge 0.5 + \triangle s$ and $\cos^2(\varphi) \le
0.5 - \triangle s$, where $\triangle s$ is the polarizer
threshold. This threshold is related to the intensity difference
between the two channels, which a PBS can resolve in the
experiments. A threshold of 0.1, for example, amounts to a
resolution better than 30\%.

In all experiments of this kind decoherence of the two separate
laser beams should play an important role. Decoherence in the
experiments has essentially two origins: (i) The random deviation
of the phase of laser beams from the phase of monochromatic waves;
and (ii) the random deviation of the location of source and
analyzers due to thermal vibration. In the experiments the
correlations were independent of distance variations
\cite{aspect81}. Therefore the laser beams are assumed to be
ideal. But, as shown above, the angle of polarization depends on
the location of the measurement. This means that we have to
include thermal vibrations in the model. This is done by a random
variation of the optical path of a photon from its source to the
analyzer. The length of the random segment is determined by the
level of decoherence. 100\% decoherence, for example, means that
half a wavelength, or a phase angle of $\pi$, is random.
Correspondingly, for 10\% decoherence a phase angle of $\pi/10$ is
random.

Numerically, the procedure is as follows. 10000 pairs of photons
are emitted from their common source. Their initial angle of
polarization is created by a random number generator
\cite{random}, the random number [0,1] is mapped onto the interval
[0,$2\pi$]. We add an angle of $\pi/2$ for one photon of each
pair. The polarization angle is set equal to the initial phase
angle $\psi_0$. Both photons cover an optical path of $L \pm
\triangle L$, where $\triangle L$ is the random segment due to
decoherence, created by a separate random input. After covering
the path to the analyzers, both photons are measured. The
variables in this measurement are the phase angle $\psi$ (= the
hidden variable), and a polarizer angle $\alpha$. We formalize the
actual measurement by a switch $\cos^2(\psi - \alpha) \ge 0.5 +
\triangle s$ (+), and $\cos^2(\psi - \alpha) \le 0.5 - \triangle
s$ (-). The setting of polarizer 2 remains unchanged during the
whole simulation. After 10000 pairs have been recorded, we change
the angle $\alpha$ by $\pi/100$. Compared to the experimental
practice, where the PBS remains unchanged, and only the speed of
the optical modulator is altered \cite{weihs98}, the only
difference should be a phase shift of the correlation function. In
all figures we report only the coincidences $N_{++}$.

\begin{figure}
\begin{center}
\epsfxsize=0.9\hsize \epsfbox{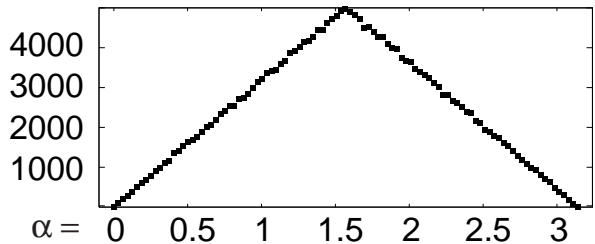}
\end{center}
\caption{
         Coincidence ($N_{++}$) counts of 10000 emitted photon
         pairs. The ideal measurement (visibility 1.0) has the
         shape of a saw-tooth.
         }
\label{fig001}
\end{figure}

Initially we simulated an ideal measurement. The fields of the two
laser beams are fully coherent throughout the distance between the
two polarizers, and the experimental devices are supposed to have
ideal characteristics. The result of this simulation is shown in
Fig. \ref{fig001}. The simulation differs from the theoretical
curve obtained by Eq. (\ref{eq003}). This difference is due to the
digital measurements, on which our model is based. While Furry's
integral \cite{furry36} predicts a minimum of $N/8$ for the
correlation curve, where $N$ is the number of photon pairs, we
obtain zero. The visibility $(N_{max} - N_{min})/(N_{max} +
N_{min})$ for an ideal measurement is therefore 1, contrary to
0.5, obtained by the integration. The simulation also differs from
the prediction in quantum mechanics, which is described by $N/2
\sin^2(\alpha - \beta)$. The correlation curve in the simulation
has a different functional form: while it is sinusoidal in quantum
mechanics, it is a saw tooth in our model. This difference is due
to the omission of interference terms, or the disregard for
superposition. The maximum ($N/2$) and the minimum (zero),
however, are equal in both cases. We also simulated the effect of
the polarizer threshold $\triangle s$ on the result of ideal
measurements. The threshold $\triangle s$ was varied from 0.0 to
0.2. Here, we find that the threshold reduces the absolute yield
of coincidences in a simulation and increases the width of the
minimum at the angles 0 and $\pi$. This effect is equal to a
retarded onset of the correlation function at its minimum
position.

\begin{figure}
\begin{center}
\epsfxsize=0.9\hsize \epsfbox{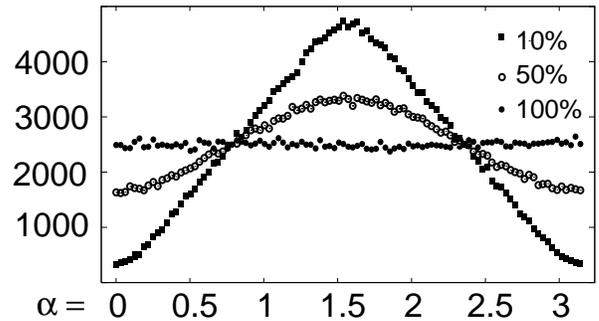}
\end{center}
\caption{
         Coincidence counts for non-ideal measurements.
         Decoherence renders the resulting curves (10\%, 50\%)
         sinusoidal, the fully decoherent curve (100\%) shows
         that correlations in our model are only due to the connecting
         phase.
        }
\label{fig002}
\end{figure}

The polarizers in current measurements are more than 400 m apart
\cite{weihs98}. Furthermore, there is no vibration damping or
cooling to very low temperatures involved in such a measurement.
This feature of the measurements is bound to cause random motion
of system components. From surface science the range of motion
without damping can be estimated, it should be for an isolated
surface no less than a few nanometers or more than one percent of
the photon's wavelength. Considering that we deal with three
coupled components and optical paths in between it seems safe to
increase this estimate by one order of magnitude. In this case we
have to include random motion of our system in the range of about
5-10\% of the wavelength. This translates, in our model, into a
decoherence rate of 10-20\% (100\% means that half a wavelength of
the photon's optical path is random). Simulations with a
decoherence rate of 10\%, 50\% and 100\% are shown in Fig.
\ref{fig002}. The interesting feature of decoherence is that it
renders the resulting distribution more sinusoidal than the
correlation function of an ideal measurement. The fully decoherent
simulation proves that the correlation, in our model, depends only
on a phase connection between the two photons.

If decoherence due to thermal vibrations of the system components
is assumed to be about 10\%,we may estimate the actual threshold
$\triangle s$ in the experiments from the experimentally obtained
visibility. In the 1998 experiment the visibility was reported to
be 97\% \cite{weihs98}. This is not in keeping with an ideal
combination PBS/photon detectors, since a threshold $\triangle s =
0.0$, under the condition of 10\% decoherence, leads to a
visibility of only 86\%. This value is also lower than the
visibility of $\approx$ 88\%, reported in the 1981 experiment
\cite{aspect81}. From the experimental facts we have to conclude
that the threshold $\triangle s$ cannot be zero. This conclusion
is also consistent with previous considerations, which established
that a finite threshold is necessary to decide on the outcome of a
polarization measurement using a PBS. Then the only remaining
question is the actual value of $\triangle s$. We only obtain a
visibility of close to 97\%, as in the 1998 experiments, if
$\triangle s \approx 0.1$.

\begin{figure}
\begin{center}
\epsfxsize=1.0\hsize \epsfbox{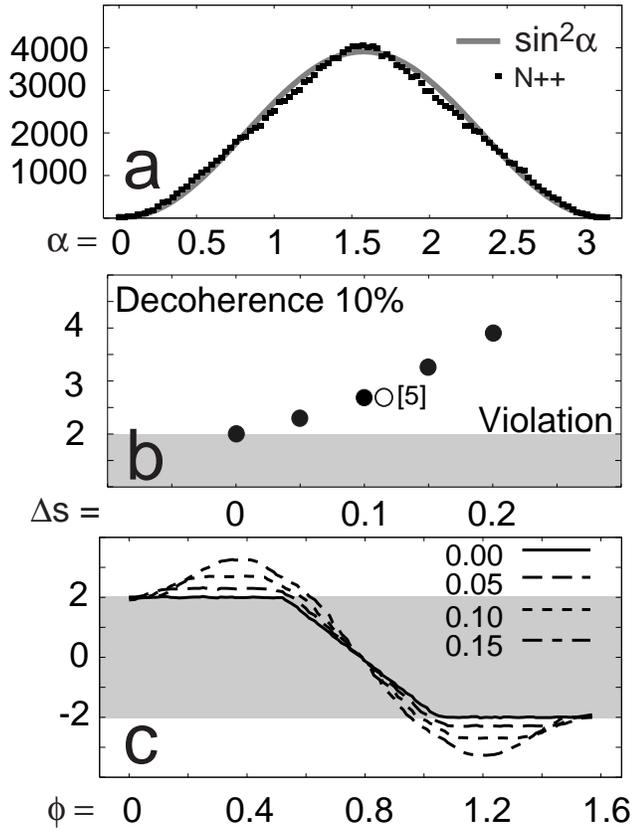}
\end{center}
\caption{
         (a) Correlation curve for the most likely experimental
         parameters. (b) CHSH value [20] for
         varying threshold $\triangle s$ and fixed decoherence of
         10\%. The Bell inequalities are violated in all cases
         where $\triangle s$ is not zero. (c) S($\phi$)
         [21] for $0 < \phi < \pi/2$ and varying
         $\triangle s$. The curve for $\triangle s = 0.1$ is equal
         to the experimental results.
         }
\label{fig003}
\end{figure}

In Fig. \ref{fig003} (a) we show the result of our simulation
under these conditions. It can be seen that the correlation curve
in this case is very similar to the ideal curves from quantum
mechanical models, it is also virtually identical to the curves
obtained in the 1998 experiment \cite{weihs98}. It has been
claimed that these experiments rule out "objective local theories"
or "realistic local theories"
\cite{clauser74,clauser78,zeilinger99}. The basis of this claim
was the assertion that within both classes of theories the Bell
inequalities are not violated in such an experiment. To show that
our model violates the Bell inequalities, we simulate the counts
at four selected angular positions of the polarizers $\alpha$ and
$\beta$ (0, 45, 22.5, 67.5). These positions yield the maximum
violation of Bell's inequalities in the standard framework. We
performed the simulations for varying threshold values from 0.0 to
0.2, and for a fixed decoherence rate of 10\%. For every setting
we performed 10 separate runs, each with 10000 pairs of photons.
Fig. \ref{fig003} (b) gives the result of our simulation. The
violation was computed according to the version of Clauser {\it et
al.} (CHSH) \cite{clauser69}. As shown, it increases with
increasing threshold. Furthermore, the limit of violation is close
to 2.0 (CHSH value of 3.90) in the final setting. For a threshold
of 0.1 and a decoherence rate of 10\%, which we found close to the
experimental conditions, the CHSH value is 2.69 $\pm$ 0.02, as
compared to 2.73 $\pm$ 0.02 reported in the 1998 experiments
\cite{weihs98}. The results are therefore in excellent agreement.
Finally, we simulated Bell violations over the whole range of
polarizer angles, using the formulation employed in Aspect's 1982
experiments \cite{phi}. Also in this simulation decoherence was
10\%, the number of photon pairs was 10000. The result of this
simulation is shown in Fig. \ref{fig003} (c). It can be seen that
it is virtually identical, for realistic conditions, to the result
obtained in the experiment \cite{aspect82}.

Given the fact that our model is a local and realistic model of
the experiments, the claim that these EPR experiments rule out any
"realistic local theory" is only valid, if the experiments are in
fact ideal. Which we think they are not. But instead of using
current experimental results as arguments how much or how little
the various no-go theorems apply, we shall suggest controlled
experiments in future publications. There, the non-ideality can be
related to model parameters, and these in turn allow extrapolating
the experimental results to ideal situations \cite{hofer01}. We
think that this procedure may ultimately provide a better
understanding of the experiments than exists at present.

A referee has asked whether the author actually ''believes'' that
this model ''describes nature''. I think this is not a scientific
question. Every model, which correctly describes an experiment,
describes nature. The only relevant question is whether one model
has advantages compared to another one. Given that this model is
local, it is realistic, and it explains the observed correlation
in a simple manner, it has at least three advantages compared to
the standard model in quantum mechanics. Whether it is correct on
a more general level, depends on the results of subsequent
experimental research.

Helpful discussions with G. Adenier, D. Bowler, A. Fisher, J.
Gavartin, J. Gittings, and R. Stadler are gratefully acknowledged.
Computing facilities at the UCL HiPerSPACE center were funded by
the Higher Education Funding Council for England.


%

\end{document}